\documentclass{iopart}
\usepackage{graphicx}
\usepackage{amsfonts}
\usepackage{amssymb}
\usepackage{psfrag}

\def\be{\begin{equation}}
\def\ee{\end{equation}}
\def\ba{\begin{eqnarray}}
\def\ea{\end{eqnarray}}

\def\ket#1{|{#1}\rangle}
\def\bra#1{\langle{#1}|}
\def\braket#1#2{\langle{#1}|{#2}\rangle}

\def\identite{\mathbb{I}}

\begin{document}
\title[Improved MPO Renormalization Group]
      {Improved Matrix Product Operator Renormalization Group: application to
      the $N$-color random Ashkin-Teller chain}

\author{Christophe Chatelain}
\address{Universit\'e de Lorraine, CNRS, LPCT, F-54000 Nancy, France}
\ead{christophe.chatelain@univ-lorraine.fr}
\date{\today}

\begin{abstract}
 Strong-Disorder Renormalization Group (SDRG), despite being a relatively simple real-space renormalization procedure, provides in principle exact results on the critical properties at the infinite-randomness fixed point of random quantum spin chains. Numerically, SDRG can be efficiently implemented as a renormalization of Matrix Product Operators (MPO-RG). By considering larger blocks than SDRG, MPO-RG was recently used to compute non-critical quantities of finite chains that are inaccessible to SDRG. In this work, the accuracy of this approach is studied and two simple and fast improvements are proposed. The accuracy on the ground state energy is improved by a factor at least equal to 4 for the random Ising chain in a transverse field. Finally, the proposed algorithms are shown to yield Binder cumulants of the 3-color random Ashkin-Teller chain that are compatible with a second-order phase transition while a first-order one is predicted by the original MPO-RG algorithm.
\end{abstract}


\section{Introduction}
The critical behavior of the random quantum Ising chain in a transverse
field (RIMTF) is known to be governed by a very peculiar renormalization-group
fixed point where randomness becomes infinitely strong~\cite{Fisher1,Fisher2,Igloi1}.
The properties of this Infinite-Disorder quantum critical point were elucidated
using a relatively simple real-space renormalization group,
previously introduced by Ma and Dasgupta~\cite{Ma,Dasgupta},
and known as Strong-Disorder
Renormalization Group (SDRG)~\cite{Monthus1,Monthus2}. The term $H_0$ of the
Hamiltonian with the largest coupling is isolated from the rest of the chain.
The full Hilbert space of the spin chain is then projected out onto the
subspace spanned by the ground states of $H_0$. A strong transverse field
$h_i$ leads to a freezing of the spin on which it acts while a strong exchange
coupling $J_i$ freezes the relative states of the two spins at its edges.
The latter can be considered as a two-state effective macro-spin.
Effective interactions with the rest of the chain are generated by
second-order perturbation theory. An effective exchange coupling
$J_{\rm eff}=J_{i-1}J_i/h_i$ is induced between the two neighboring
spins of a spin frozen by a strong transverse field $h_i$. Similarly,
an effective transverse field $h_{\rm eff}=h_{i}h_{i+1}/J_i$ acts on the
macro-spin formed by a strong exchange coupling. As the renormalization
is iterated, the probability distribution of the couplings evolves towards
an infinitely broad law. As a consequence, a strong coupling is more and
more likely to be surrounded by weak couplings. Therefore, the SDRG is
believed to become exact, not only at the IRFP but in the whole Griffiths
phase~\cite{Igloi}.
\\

Following the general principles of renormalization group, the critical
exponents are extracted from the flow equations of couplings during
the renormalization process. The dynamical exponent $z$ for instance
is obtained from the scaling of the number of remaining sites while
the magnetic exponent $\beta$ is given by the scaling of the total
magnetic moment of the chain. In the case of the random Ising chain in a
transverse field, the flow equations have been solved by Fischer. For more
general models, as for instance the random Ashkin-Teller model, these equations
cannot be solved but SDRG rules can easily be implemented numerically~\cite{Hoyos1,
Hoyos2,Hoyos3}. Even though very approximate effective interactions are generated
during the first iterations of the SDRG, they are expected to become more and
more accurate as the IRFP is approached. It is therefore necessary to
apply the technique to very large chains, typically of the order of
tens of thousands or millions of spins. The procedure is nevertheless able
to give accurate estimates of critical exponents. Moreover, SDRG can be
implemented numerically to study lattice models in higher dimensions~\cite{Igloi2}.
\smallskip

For strong disorder, SDRG is the most efficient technique to estimate numerically
the critical exponents. The Density Matrix Renormalization Group (DMRG) algorithm~\cite{White1,White2,Schollwoeck1,Schollwoeck2} suffers
from stringent convergence problems in presence of strong disorder. In the
case of the above-mentioned random Ashkin-Teller model for example, only small
lattices could be considered~\cite{Carlon,Chatelain}. However, SDRG allows for
numerical estimates of the critical exponents but not of the quantum averages
at any point of the phase diagram. MPO renormalization, as introduced in Refs~\cite{Hikihara,Goldsborough} and then considered in~\cite{Lin}, is an attempt
to fill the gap between DMRG and SDRG. As in DMRG, an effective Hamiltonian acting
on a small Hilbert space is iteratively constructed and quantum averages are estimated
in the ground state of this Hamiltonian. However, in contrast to DMRG and as SDRG,
the technique is more efficient at strong disorder. MPO renormalization is therefore
meant as an alternative to DMRG at strong disorder rather than an extension of SDRG.
Like the Hamiltonian, the observables should be expressed as MPO.
Powers of global observables $(\sum_i O_i)^n$ can also be written as MPO~\cite{Lin}.
At each step of the renormalization process, the same transformation is applied to the
matrix product of the Hamiltonian and of all observables. At
the end of the renormalization, i.e. when only one site remains, the
Hamiltonian is diagonalized and the averages of the observables are
computed in the ground state. In the case of the random anti-ferromagnetic
Ising chain in a transverse field, the Binder cumulant was estimated with this
algorithm and the location of its crossing points were shown to be in good
agreement with the exact transition point~\cite{Lin}.
\\

In this study, two improvements of the MPO renormalization algorithm are
introduced. They are tested in the case of the random Ising chain in a
transverse field and then used to determine the phase diagram of the 2 and 3-color
Ashkin-Teller model.
In the first section of this paper, SDRG is reviewed. The emphasis is put
on the construction of effective interactions by perturbation theory. In
the second section, the MPO renormalization algorithm is presented. The
equivalence with SDRG in the limit of strong couplings is shown in the
particular case of the Ising chain in a transverse field. In the third section,
our improvements of this algorithm are presented: a new criterion is introduced
to choose the blocks to be merged in the renormalization procedure and the
construction of effective interactions taking into account the highest
excited states to be discarded is presented. In the fourth section, the accuracy
of the estimates of the average ground state energy and of the gap with the
first excited state of these two algorithms is compared with the original MPO-RG.
The method is also applied to compute the Binder cumulant of the random
Ising chain in a transverse field. In the last section, the algorithm is applied
to the 2 and 3-color Ashkin-Teller model. Conclusions follow.

\section{Review of Strong-Disorder Renormalization rules}
Consider the random Ising chain in a transverse field whose Hamiltonian reads
   \be H=-\sum_{i=1}^{N-1} J_i\sigma_i^x\sigma_{i+1}^x-\sum_{i=1}^N h_i\sigma_i^z
   \label{Ising}\ee
where the couplings $J_i$ and $h_i$ are random variables. The SDRG algorithm is
the following: find the strongest coupling $\Omega={\rm max}_i\{J_i,h_i\}$.
Isolate the term $H_0$ of $H$ involving $\Omega$.
Restrict the Hilbert space to the subspace spanned by the ground states
of $H_0$. Generate effective interactions with the rest of the chain
using second-order perturbation theory. Iterate until leaving only one site.

In the case of $\Omega=h_i$ for example, the local Hamiltonian on site $i$ is
   \be H_0=-h_i\sigma_i^z\ee
so the ground state is $\ket\uparrow_i$ (if $h_i>0$). The Hilbert space
is projected out onto the subspace spanned by $\{\ket\uparrow_i\}$ with
the projection operator
   \be P=\ket\uparrow_i\bra\uparrow_i
   =\identite^{\otimes i-1}\otimes\ket\uparrow\bra\uparrow
   \otimes \identite^{\otimes N-i}.\ee
As a result, the spin is frozen in the state $\ket\uparrow_i$. An effective coupling
between the spins $i-1$ and $i+1$ is computed with the perturbing Hamiltonian
 \be W=-J_{i-1}\sigma_{i-1}^x\sigma_i^x-J_i\sigma_i^x\sigma_{i+1}^x.\ee
It is convenient to consider the Dyson expansion of the perturbed Green function
    \ba\fl(z-W_{\rm eff})^{-1}=PG(z)P
    &=&{1\over z-H_0}+{1\over z-H_0}PWP{1\over z-H_0}\\
    &&\quad\quad +{1\over z-H_0}PW{1\over z-H_0}WP{1\over z-H_0}+\ldots\nonumber\ea
The first order term of the matrix element ${\bra\uparrow}_iG(z){\ket\uparrow}_i$
vanishes and, since $\sigma_i^x\ket{\uparrow}_i=\ket{\downarrow}_i$,
    \be\fl{\bra\uparrow}_iG(z){\ket\uparrow}_i
      ={1\over z+h_i}+{1\over z+h_i}\big[J_{i-1}\sigma_{i-1}^x
      +J_i\sigma_{i+1}^x\big]{1\over z-h_i}
      \big[J_{i-1}\sigma_{i-1}^x+J_i\sigma_{i+1}^x\big]{1\over z+h_i}.\ee
Note that $1/(z-h_i)$ is the unperturbed Green function evaluated in the
excited state. Since we are interested in an effective interaction in the ground state,
the parameter $z$ of this unperturbed Green function is set to $z=-h_i$:
   \be{\bra\uparrow}_iG(z){\ket\uparrow}_i
    ={1\over z+h_i}-{1\over (z+h_i)^2}{J_{i-1}^2+J_{i+1}^2
    +2J_{i-1}J_i\sigma_{i-1}^x\sigma_{i+1}^x\over 2h_i}.\ee
The last term can be interpreted as a first-order term
$G_0(z) W_{\rm eff}G_0(z)$ for the effective Hamiltonian
    \ba W_{\rm eff}&=&-{J_{i-1}^2+J_{i+1}^2+2J_{i-1}J_i
    \sigma_{i-1}^x\sigma_{i+1}^x\over 2h_i}\nonumber\\
    &=&{\rm Cste}-{J_{i-1}J_i\over h_i}\sigma_{i-1}^x\sigma_{i+1}^x
    \ea
i.e. an effective exchange coupling $J_{\rm eff}=J_{i-1}J_i/h_i$.
    
Similarly, if the strongest coupling is $J_i$, the ground states of
$H_0=-J_i\sigma_{i}^x\sigma_{i+1}^x$ are $\ket{\tilde\uparrow}_{i+1}
=\ket{\uparrow_x}_i\otimes\ket{\uparrow_x}_{i+1}$ and $\ket{\tilde\downarrow}_{i+1}
=\ket{\downarrow_x}_i\otimes\ket{\downarrow_x}_{i+1}$. The Hilbert space is
projected out onto the subspace spanned by these two states. $\tilde\sigma_{i+1}$
behaves as a macro-spin. The excited states induce an effective interaction
   \be W_{\rm eff}={\rm Cste}-{h_ih_{i+1}\over J_i}\tilde\sigma_i^z.\ee

The method becomes exact as the infinite-randomness fixed point is
approached because the probability distribution of the couplings is
broader and broader. A strong coupling is more likely to be surrounded by weak
couplings, justifying the use of perturbation theory.

\section{RG algorithms for MPO}
\subsection{MPO formulation of renormalization}\label{MPO-RG}
Consider an open spin chain of $N$ spins with the Hamiltonian
  \be H=\sum_{i=1}^N H_i+\sum_{i=1}^{N-1}L_iR_{i+1} \label{Hamilt}\ee
where $R_i=\identite^{\otimes i-1}\otimes R\otimes\identite^{\otimes N-i}$
for instance acts on the $i$-th spin.
Using successive Singular Value Decompositions (SVD), the matrix
elements of any linear operator
    \be\hat O=\sum_{\sigma_1,\ldots,\sigma_N,\atop\sigma_1',\ldots,\sigma_N'}
    O_{\sigma_1,\ldots,\sigma_N;\sigma_1',\ldots,\sigma_N'}\ket{\sigma_1,
    \ldots,\sigma_N}\bra{\sigma_1',\ldots,\sigma_N'}\ee
acting on the Hilbert space ${\cal H}_1^{\ \otimes N}$ of the $N$ spins
can be cast as a product of matrices~\cite{Verstraete,Orus1,Orus2}
    \be O_{\sigma_1,\ldots,\sigma_N;\sigma_1',\ldots,\sigma_N'}
    =(A_1)^{\sigma_1,\sigma_1'}_{\ \ a_1}(A_2)^{\sigma_2,\sigma_2'}_{\ \ a_1,a_2}
    \ldots (A_N)^{\sigma_N,\sigma_N'}_{\ \ a_{N-1}}.\ee
The lower indices correspond to an auxiliary vector space associated
to the bonds of the chain. This decomposition is referred to as Matrix
Product Operator. For the Hamiltonian (\ref{Hamilt}), the smallest
dimension of this auxiliary vector space is $\chi=3$ and the matrices
read
    \be A_i=\pmatrix{\identite & L_i & H_i \cr
     0 & 0 & R_i \cr
     0 & 0 & \identite \cr }\ee
for $1<i<N$ while at the two edges of the chain
    \be A_1=\pmatrix{\identite & L_1 & H_1 },\hskip 1cm
    A_N=\pmatrix{H_N \cr R_N \cr \identite}\ee
The simplest renormalization algorithm is as follows. The system is
divided into blocks of two spins. The local Hamiltonian of the block
spanning over the sites $i$ and $i+1$ is given by the matrix element
    \be (A_i\otimes A_{i+1})_{1,\chi}
    =H_i\otimes\identite+L_i\otimes R_{i+1}+\identite\otimes H_{i+1}.\ee
For each block, the local Hamiltonian is diagonalized
and the largest gap is found in the energy spectrum. The renormalization is
performed on the block with the largest energy gap. Its Hilbert space
is truncated to the subspace spanned by the eigenvectors whose eigenvalues
are below the gap. The local Hamiltonian, as well as all other non-zero
matrix elements of $A_i\otimes A_{i+1}$, are projected out onto this subspace.
This defines a renormalized matrix
   \be A'_i=U^+(A_i\otimes A_{i+1})U    \label{Transfo1}\ee
where $U$ is a rectangular matrix whose rows are the selected eigenvectors
of the local Hamiltonian. The transformation is not unitary. Note that
$U$ acts on the spin indices and not on the auxiliary vector space.
The matrix $A'_i$ has dimension $\chi\times\chi$, except at the left
and right edges of the chain, and keeps the same structure as the original
$A_i$'s. The process is iterated until the chain has a single site.

\subsection{Equivalence with SDRG}
Even though {\sl a priori} simpler than SDRG, this approach is actually
equivalent in the limit of strong randomness. Consider again the
Ising chain in a transverse field (\ref{Ising}). The Hamiltonian can be
cast as a MPO with the matrices
    \be A_i=\pmatrix{ \identite & -\sqrt{J_i}\sigma^x & -h_i\sigma^z\cr
      0 & 0 & \sqrt{J_{i-1}}\sigma^x \cr 0 & 0 & \identite}\ee
for $1<i<N$ and
    \be A_1=\pmatrix{ \identite & -\sqrt{J_1}\sigma^x & -h_1\sigma^z},\quad
     A_N=\pmatrix{ -h_N\sigma^z\cr \sqrt{J_{N-1}}\sigma^x \cr \identite}.\ee
Suppose that the largest gap is found for the block obtained after merging
sites $i$ and $i+1$. The local Hamiltonian of this block is then
    \be H_{i,i+1}=(A_i\otimes A_{i+1})_{1\chi}
    =-h_i\sigma^z\otimes\identite-h_{i+1}\identite\otimes\sigma^z
    -J_i\sigma^x\otimes\sigma^x\ee
whose four eigenvalues are
    \be \pm E_1=\pm\sqrt{(h_i+h_{i+1})^2+J_i^2},\hskip 1truecm
    \pm E_2=\pm\sqrt{(h_i-h_{i+1})^2+J_i^2}.\ee
Keeping the two states below the largest gap, i.e. with energies $-E_1$ and
$-E_2$, the effective matrix is
    \be\fl A_i'=U^+A_iA_{i+1}U=\pmatrix{ \identite &
     -\sqrt{J_{i+1}}U^+(\identite\otimes\sigma^x)U &
     \pmatrix{ -E_1 & 0 \cr 0 & -E_2} \cr  0 & 0 &
     \sqrt{J_{1}}U^+(\sigma^x\otimes\identite)U \cr
     0 & 0 & \identite}\ee
By construction, the renormalized local Hamiltonian is diagonal in this basis
and can therefore be written as
    \be H_{i,i+1}=-{1\over 2}(E_1+E_2)\identite-{1\over 2}(E_1-E_2)\sigma^z
    ={\rm Cst}\ \!\identite-h_{\rm eff}\sigma^z\ee
with the effective transverse field
    \be\fl h_{\rm eff}={1\over 2}(E_1-E_2)
    ={1\over 2}\sqrt{(h_i+h_{i+1})^2+J_i^2}
    -{1\over 2}\sqrt{(h_i-h_{i+1})^2+J_i^2}\ee
It turns out that the renormalized operators $U^+(\sigma^x\otimes\identite)U$
and $U^+(\identite\otimes\sigma_i^x)U$ are proportional to $\sigma^x$ so
the expression of the original Hamiltonian is preserved~\footnote{If the
rotated Hamiltonian $H=-J\sum_i \sigma_i^z\sigma_{i+1}^z
-h\sum_i \sigma_i^x$ is considered instead of (\ref{Ising}),
an additional $45^\circ$ rotation is needed at each renormalization step
to bring back the local Hamiltonian to its original form.}.
\\

\begin{figure}
  \begin{center}
    {
      \psfrag{h}[Bl][Bl][1][1]{$h_{\rm eff}$}
      \psfrag{Jlr}[Bl][Bl][1][1]{$J_{i-1}$, $J_{i+1}$}
      \psfrag{heff}[tc][tc][1][0]{$h_{\rm eff},J_{i-1},J_{i+1}$}
      \psfrag{J}[Bc][Bc][1][1]{$J_i$}
      \includegraphics[width=6.25cm]{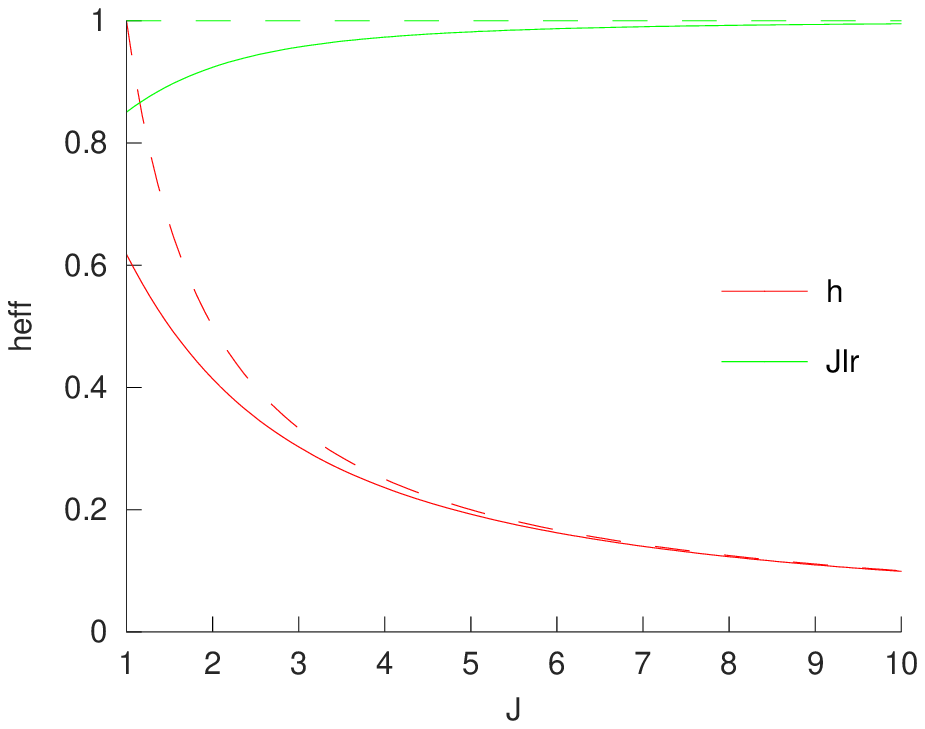}
    }{
      \psfrag{h}[Bl][Bl][1][1]{$h_i$}
      \psfrag{Jl}[Bl][Bl][1][1]{$J_{i-1}$}
      \psfrag{Jr}[Bl][Bl][1][1]{$J_{i+1}$}
      \psfrag{heff}[tc][tc][1][0]{$h_{\rm eff}$}
      \psfrag{Jeff}[Bc][Bc][1][1]{$J_{i-1},J_{i+1},h_{\rm eff}$}
      \includegraphics[width=6.25cm]{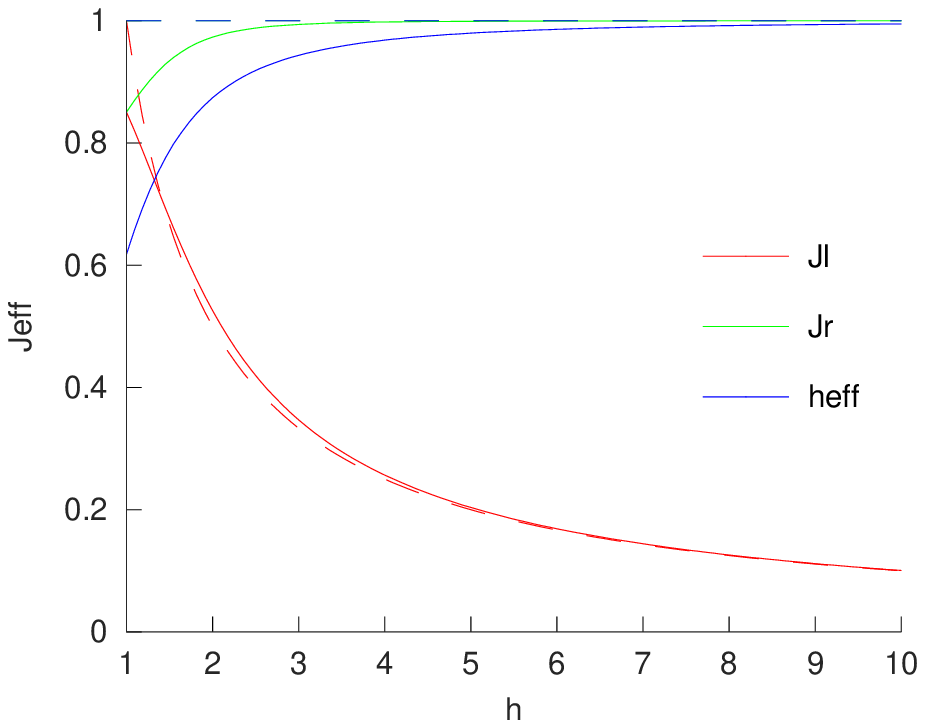}
    }
    \caption{On the left, effective couplings of a 2-spin block after renormalization
      versus the exchange coupling $J_i$ binding the two spins. All other
      couplings (transverse fields and couplings with the spins at the left and
      the right of the block) are taken equal to one. The dashed lines
      are the predictions of SDRG. Note that the latter predicts an absence
      of renormalization of the left and right couplings ($J_{i-1}$ and
      $J_{i+1}$ if the block spans over the sites $i$ and $i+1$).
      On the right, effective couplings of a 2-spin block after renormalization
      versus the transverse field $h_i$ originally coupled to the left spin
      of the block. All other couplings (transverse field and couplings inside
      and outside the block) are taken equal to one. The dashed lines
      are again the predictions of SDRG.
    }
  \label{fig1}
  \end{center}
\end{figure}

When the exchange coupling $J_i$ is stronger than both $h_i$ and $h_{i+1}$,
a Taylor expansion to lowest-order in $(h_i\pm h_{i+1})/J_i$ gives the
SDRG renormalized transverse field
     \be h_{\rm eff}\simeq {J_i\over 2}\Big(1+{(h_i+h_{i+1})^2\over 2J_i^2}\Big)
     -{J_i\over 2}\Big(1+{(h_i-h_{i+1})^2\over 2J_2^2}\Big)
     ={h_ih_{i+1}\over J_i}.\ee
When the transverse field $h_i$ is stronger than both $h_{i+1}$ and $J_i$,
the gap between the two lowest eigenvalues $-E_1$ and $-E_2$ is now
    \be {1\over 2}(E_1-E_2)
    \simeq {h_i\over 2}\Big[1+{h_{i+1}\over h_i}+{\cal O}
      \Big({1\over h_i^2}\Big)\Big]-{h_i\over 2}\Big[1-{h_{i+1}\over h_i}
      +{\cal O}\Big({1\over h_i^2}\Big)\Big]=h_{i+1}\ee
i.e. equal to the original transverse field acting on site $i+1$.
In the basis $\{\ket{\uparrow\uparrow},\ket{\downarrow\uparrow},
\ket{\uparrow\downarrow},\ket{\downarrow\downarrow}\}$,
the associated eigenvectors are proportional to
   \ba
   &&\pmatrix{ E_1+h_i+h_{i+1} & 0 & 0 & J_i }
   \simeq \pmatrix{ 2(h_i+h_{i+1}) & 0 & 0 & J_i }\nonumber\\
   &&\pmatrix{ 0 & E_2-h_i+h_{i+1} & J_i & 0 }
   \simeq \pmatrix{ 0 & {J_i^2\over 2h_i} & J_i & 0}
   \ea
to lowest-order in $1/h_i$. The 2-spin block is coupled to $\sigma_{i-1}$
via the operator $\sigma_i^x$. After renormalization, $\sigma_i^x$ is
transformed into $U^+(\sigma^x\otimes\identite)U$. The latter is proportional
to $\sigma^x$. The coefficient is computed as the off-diagonal
matrix element between the two (normalized) eigenvectors
   \ba &&{1\over 2h_i}\pmatrix{ 2h_i & 0 & 0 & J_i }\pmatrix{
     0  & 1 &  0 &  0 \cr
     1  & 0 &  0 &  0 \cr
     0  & 0 &  0 &  1 \cr
     0  & 0 &  1 &  0 }\ \!
   {1\over J_i}\pmatrix{ 0 \cr {J_i^2\over 2h_i} \cr J_i \cr 0 \cr}
   ={J_i\over h_i}\ea
The Hamiltonian coupling $\sigma_{i-1}$ and the 2-spin block is
therefore
   \be -J_{i-1}\sigma_{i-1}^xU^+(\sigma^x\otimes\identite)U
   =-{J_{i-1}J_i\over h_i}\sigma_{i-1}^x\tilde\sigma_{i+1}^x\ee
as predicted par SDRG. A comparison of the renormalized couplings
as estimated by SDRG and MPO-RG is shown on figures~\ref{fig1}.

\section{Improvements of the MPO renormalization algorithm}
\subsection{New criterion for selecting the block to be renormalized}\label{MPO-RG2}
In the above-described renormalization algorithm, a low-energy effective
Hamiltonian is constructed by successive projections onto the lowest
eigenstates of local Hamiltonians. At each iteration, the two-spin block
to be renormalized is therefore treated as completely decoupled from the
rest of the chain. Close to the IRFP, randomness becomes very large so,
if one of the inter-block couplings is strong, one can safely assume that 
the couplings with the rest of the chain are much smaller.
Away from the IRFP, this is no more the case and the interaction with
the rest of the chain cannot be neglected. The renormalization procedure
is then expected to introduce systematic deviations on the ground state
of the whole chain.
\\

To partially circumvent the problem, a simple approach consists in
renormalizing in priority the block with, not only the largest gap in
the spectrum of its local Hamiltonian, but also with the smallest couplings
with the rest of the chain. We suggest the following modification to the
algorithm: the ground state energy $\varepsilon_0^{(i,i+1)}$ is first computed
for each block of two sites $(i,i+1)$. The strength of the coupling between
the two spins is estimated as the difference
   \be\Delta\varepsilon_0^{(i,i+1)}=\varepsilon_0^{(i)}
   +\varepsilon_0^{(i+1)}-\varepsilon_0^{(i,i+1)}\ee
where $\varepsilon_0^{(i)}$ is the energy of the single spin at site $i$.
Then, to compare the inter-block coupling with the couplings of the two
neighboring blocks, the ratio
   \be \rho^{(i,i+1)}={\Delta\varepsilon_0^{(i,i+1)}\over
     {\rm max}\ \!\big(\Delta\varepsilon_0^{(i-1,i)},
     \Delta\varepsilon_0^{(i+1,i+2)}\big)}\ee
is computed for each block. Last, the renormalization is performed on the
block with the largest ratio $\rho^{(i,i+1)}$. This simple modification is
observed to give lower ground state energies, closer to the estimate of
DMRG. Note that the energies $\varepsilon_0^{(i,i+1)}$ and the ratios
$\rho^{(i,i+1)}$ do not need to be computed at each renormalization step.
Only the two of them that are affected by the renormalization of a block
needs to be recomputed. Moreover, the ratio $\rho^{(i,i+1)}$ can be stored
in a binary tree in order to speed up the search for the largest one.

\subsection{Effective interactions between effective spins}\label{MPO-RG3}
A second improvement consists in generating the effective interactions
mediated by the highest eigenstates between a block and its neighboring spins.
The algorithm is as follows. A two-spin block, say $(i,i+1)$ is chosen
according to the above-described criterion. A new macro-spin is defined
by merging the two spins $i$ and $i+1$. Its local Hamiltonian $H_{i,i+1}
=(A_i\otimes A_{i+1})_{1,\chi}$ is diagonalized:
   \be H_{i,i+1}=\sum_{j=0}^{d_id_{i+1}-1} \varepsilon_j^{(i,i+1)}
   \ket{\phi_j}\bra{\phi_j}.\ee
The Hamiltonian of the macro-spin, including the interaction with its
two neighbors, is
   \be H=L_{i-1}R_i+H_{i,i+1}+L_{i+1}R_{i+2}\ee
Define the projectors
   \ba P&=&\sum_{j\le\Lambda} \identite^{\otimes i-1}\otimes\ket{\phi_j}
   \bra{\phi_j}\otimes\identite^{\otimes N-i-1},\nonumber\\
   Q&=&\sum_{j>\Lambda} \identite^{\otimes i-1}\otimes
   \ket{\phi_j}\bra{\phi_j}\otimes\identite^{\otimes N-i-1}
   =\identite^{\otimes N}-P\ea
where the cut-off $\Lambda$ separates the eigenstates to be kept from
those to be discarded. In the original MPO renormalization-group
algorithm, the Hamiltonian is projected out onto the subspace spanned
by the lowest eigenstates, i.e. $H$ is replaced by
   \be PHP=L_{i-1}PR_iP+PH_{i,i+1}P+PL_{i+1}PR_{i+2}.\ee
To take into account perturbatively the highest eigenstates, one can
decompose the Hamiltonian as $H=H_0+W$ where the unperturbed Hamiltonian
    \be H_0=L_{i-1}PR_iP+H_{i,i+1}+PL_{i+1}PR_{i+2}\ee
does not couple the lowest and highest eigenstates and the perturbation
reads
    \be\fl W=L_{i-1}\big(PR_iQ+QR_iP+QR_iQ\big)
    +\big(PL_{i+1}Q+QL_{i+1}P+QL_{i+1}Q\big)R_{i+2}.\ee
The Dyson expansion of the perturbed Green function is
    \ba PG(z)P&=&P(z-H_0-W)^{-1}P\nonumber\\
   &=&P\big[\identite-(z-H_0)^{-1}W\big]^{-1}(z-H_0)^{-1}P\nonumber\\
   &=&\sum_{n=0}^{+\infty} P\big[G_0(z)W\big]^nG_0(z)P\nonumber\\
   \ea
where $G_0(z)=(z-H_0)^{-1}$ is the unperturbed Green function.
The first-order term vanishes because $[P,G_0]=0$ and $PWP=0$.
At second order, the Dyson expansion is
    \ba PG(z)P&=&G_0(z)+G_0(z)PWG_0(z)WPG_0(z)\nonumber\\
    &=&G_0(z)+G_0(z)\Sigma_{\rm eff}G_0(z)\ea
with the self-energy
    \be \Sigma_{\rm eff}(z)=PWG_0(z)WP.\ee
Note that $PWP=0$ so the latter can be written
    \be \Sigma_{\rm eff}(z)=PWQG_0(z)QWP.\ee
Since we are interested in the ground state of the chain, a low-energy
effective Hamiltonian is $W_{\rm eff}=\Sigma_{\rm eff}(z)$ where $z$
should be chosen equal to the ground state energy of the chain.
Different interactions are generated:
    \be\fl L_{i-1}^2PR_iQG_0(z)QR_iP
    +R_{i+2}^2PL_iQG_0(z)QL_{i+1}P
    =L_{i-1}^2X_i+Y_iR_{i+2}^2
    \label{EffectivH1}\ee
that couple the macro-spin with the spins on sites $i-1$ and $i+1$.
A three-spin interaction
    \be\fl L_{i-1}\big[PR_iQG_0(z)QL_iP
      +PL_iQG_0(z)QR_iP\big]R_{i+2}
    =L_{i-1}Z_iR_{i+2}\ee
is also generated. Taking into account these terms requires to increase
the dimension $\chi$ of the auxiliary vector space of the
matrices $A_{i-1}$, $A_i$, $A_{i+1}$. The matrices $A_{i-1}$, $A_i$,
and $A_{i+2}$ become, after renormalization,
  \be A_{i-1}=\pmatrix{ \identite & L_{i-1} & L_{i-1}^2 & H_{i-1}\cr
    0 & 0 & 0 & R_{i-1} \cr
    0 & 0 & 0 & \identite \cr },\ee
  \be A_i=\pmatrix{ \identite & PL_iP & Y_i & 0 & PH_{i,i+1}P \cr
    0 & 0 & 0 & Z_i & PR_iP \cr
    0 & 0 & 0 & 0 & X_i \cr
    0 & 0 & 0 & 0 & \identite},\ee
  \be A_{i+2}=\pmatrix{ \identite & L_{i+2} & H_{i+2} \cr
    0 & 0 & R_{i+2} \cr
    0 & 0 & R_{i+2}^2 \cr
    0 & 0 & R_{i+2} \cr
    0 & 0 & \identite}\ee   
  where
  \ba X_i(z)&=&PR_iQG_0(z)QR_iP,\nonumber\\
  Y_i(z)&=&PL_iQG_0(z)QL_iP,\nonumber\\
  Z_i(z)&=&PR_iQG_0(z)QL_iP
  +PL_iQG_0(z)QR_iP.
  \label{EffectiveInt}
  \ea
The procedure is iterated. If the sites $i$ and $i+2$ are later merged
for example, $A_{i-1}$ will be replaced by a $6\times 4$ matrix.
\\

The numerical calculation of the matrix element $\bra{\phi_k}X_i
\ket{\phi_j}$ ($k,j\le\Lambda$) has been performed in the following way: first,
$R_i$ is applied onto the eigenvector $\ket{\phi_j}$ of the local Hamiltonian.
The resulting vector is then projected out onto the levels to be discarded:
   \be \ket{\varphi}=QR_i\ket{\phi_j}=\Big[\identite
   -\sum_{k\le\Lambda} \ket{\phi_k}\bra{\phi_k}\Big]R_i\ket{\phi_j}.\ee
The unperturbed Green function $G_0(z)$ is estimated by first finding the
eigenvectors $\ket{\psi_i}$ associated to the eigenvalues $e_i$ of smallest
algebraic magnitude of the operator $z-H_0$. The numerical
calculation was performed using the implicit restarted Arnoldi algorithm as
implemented in the {\tt arpack} library. $G_0(z)QR_i\ket{\phi_j}$ is estimated
as
   \be\ket{\varphi'}=\sum_j e_j^{-1}\ket{\psi_j}\braket{\psi_j}{\varphi}\ee
The estimate is refined using a conjugate gradient algorithm.
Finally, since $G_0(z)$ is diagonal in the unperturbed basis, we do need
to apply the projector $Q$ again. The matrix element $\bra{\phi_k}X_i
\ket{\phi_j}$ is finally given by $\braket{\phi_k}{\varphi'}$.

\section{Accuracy and efficiency of the different algorithms for the random Ising chain}
In the following, the accuracy of the different approaches discussed above
is studied. Three versions of the MPO-RG algorithm are compared: the first
is the original one introduced in section~\ref{MPO-RG}, the second
implements the improved choice of the block to be renormalized of
section~\ref{MPO-RG2} and the third takes into account effective interactions
as discussed in section~\ref{MPO-RG3}. In the following, these tree variants
of the MPO-RG algorithm will be referred to as Algo 1,2, and 3.
The parameter $z$ of Algo 3 is set to the estimate of the ground state energy
given by Algo 2. For simplicity, the three-site effective interaction
(operator $Z$ in (\ref{EffectiveInt})) was neglected. The latter indeed
introduces 4-site, 5-site, $\ldots$ effective interactions as the
renormalization procedure is iterated. In contrast, the two-site effective
interactions (operators $X$ and $Y$) keep the same form during the
renormalization. We allowed for a maximum of 8 different interactions
between neighboring blocks and neglected any further interaction that would
be generated by the renormalization process. The accuracy of the different
MPO-RG algorithms is tested by comparing the estimated ground state energies.
The latter is easily computed at the end of the renormalization when only
one site is left. 

\subsection{Shift of the ground state energy during the renormalization}
To monitor the shift of the ground state energy induced by the renormalization,
the different MPO-RG algorithms were coupled to a DMRG algorithm. After each
renormalization step, a full DMRG calculation is performed on the renormalized
MPO to estimate the ground state energy. The code is drastically slowned down by
the DMRG calculations so the lattice was limited to 32 sites.
The random Ising chain in a transverse field is considered: 
    \be H=-\sum_{i=1}^{L-1} J_i\sigma_i^z\sigma_{i+1}^z-\sum_{i=1}^L h_i\sigma_i^x
   -B\sum_{i=1}^L\sigma_i^z\ee 
with a uniform probability distribution of exchange couplings $(J_i\in [0.5;2])$.
The transverse fields were also uniformly distributed but in different intervals
corresponding to different regions of the phase diagram: ferromagnetic phase
($h_i\in [0.3;0.4]$), ordered Griffiths phase ($h_i\in [0.5;1]$), critical point
($h_i\in [0.5;2]$), disordered Griffiths phase ($h_i\in [1;2]$), and
paramagnetic phase ($h_i\in [2.5;3]$). This disorder is relatively weak so
we expect the original SDRG algorithm to lead to important deviations for
small chains. On the other hand, the DMRG algorithm, used to probe these deviations,
is more efficient at weak disorder. A small longitudinal field $B=10^{-4}$
is added to further improve the convergence of the DMRG algorithm.
128 states were kept in the left and right blocks (64 for the environment
and 2 for the central spin) in the DMRG algorithm and 16 sweeps were performed.
For the three algorithms, the renormalization consisted in merging two neighboring
2-state blocks and truncating the Hilbert state to the subspace spanned by the
two eigenstates with lower energies. Results with more states per block will be
considered in the next section. Finally, the ground state energy is averaged
over 32 disorder realizations in order to show that the results are typical
and not due to a particular disorder configuration.

\begin{figure}
  \begin{center}
    \psfrag{L}[tc][tc][1][0]{$L$}
    \psfrag{Eo}[Bc][Bc][1][1]{$E_0$}
    \includegraphics[width=6.25cm]{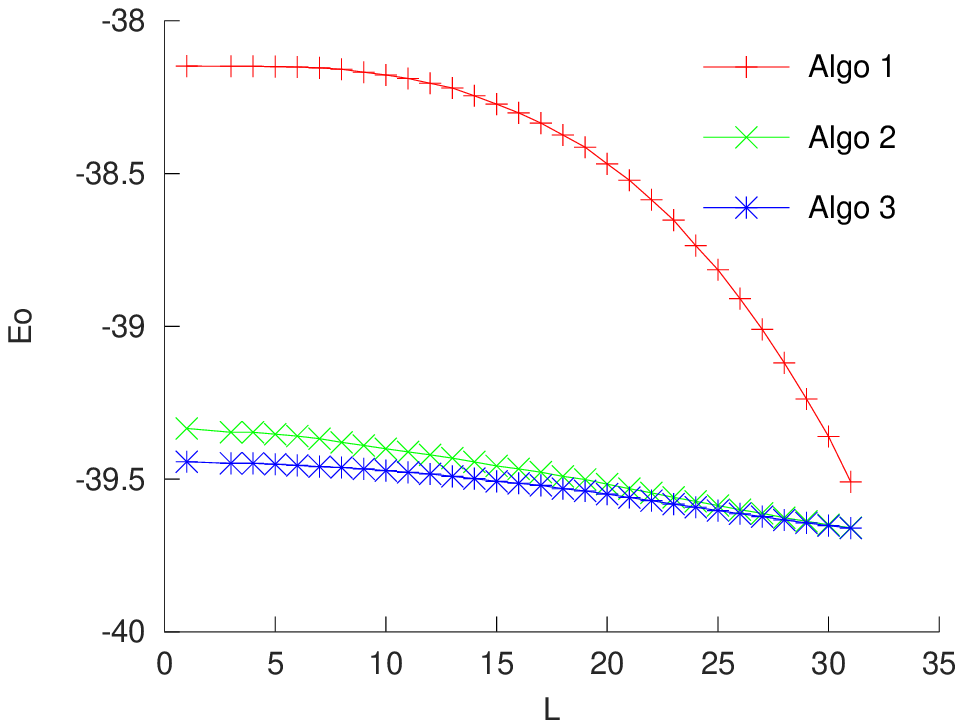}
    \includegraphics[width=6.25cm]{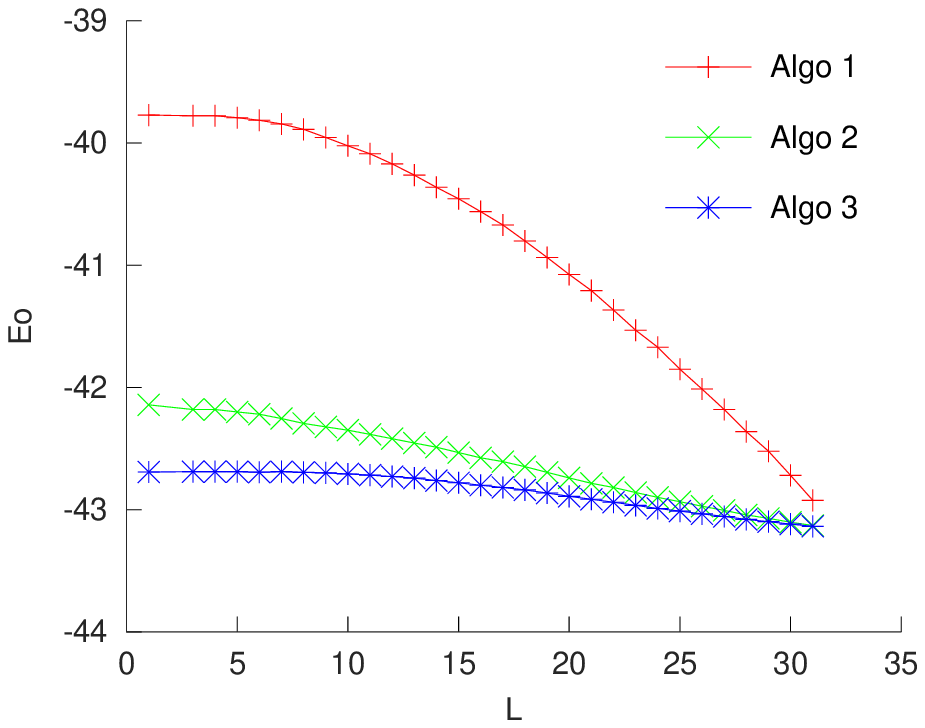}
    \caption{Average ground state energy of a random Ising chain of 32 spins
      as estimated by the three variants of the MPO-RG algorithm
      versus the number of remaining sites $L$ during the RG process.
      The system is in the ferromagnetic phase ($h_i\in [0.3;0.4]$) on
      the left figure and in the ordered Griffiths phase ($h_i\in [0.5;1]$) 
      on the right.
    }
  \label{fig3}
  \end{center}
\end{figure}

\begin{figure}
  \begin{center}
    \psfrag{L}[tc][tc][1][0]{$L$}
    \psfrag{Eo}[Bc][Bc][1][1]{$E_0$}
    \includegraphics[width=8cm]{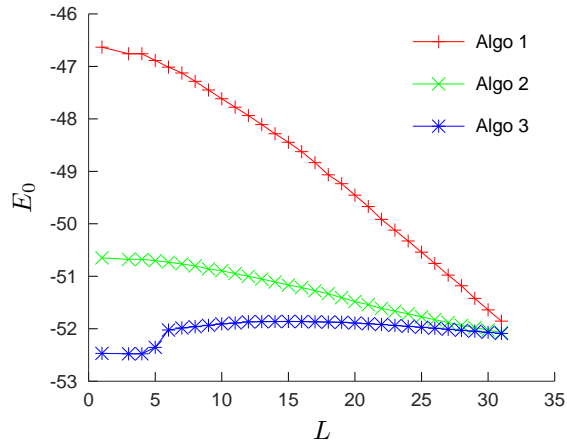}
    \caption{Average ground state energy  of a random Ising chain of 32 spins
      as estimated by the three variants of the MPO-RG algorithm
      versus the number of remaining sites $L$ during the RG process.
      The system is at the critical point ($h_i\in [0.5;2]$).
    }
  \label{fig5}
  \end{center}
\end{figure}

\begin{figure}
  \begin{center}
    \psfrag{L}[tc][tc][1][0]{$L$}
    \psfrag{Eo}[Bc][Bc][1][1]{$E_0$}
    \includegraphics[width=6.25cm]{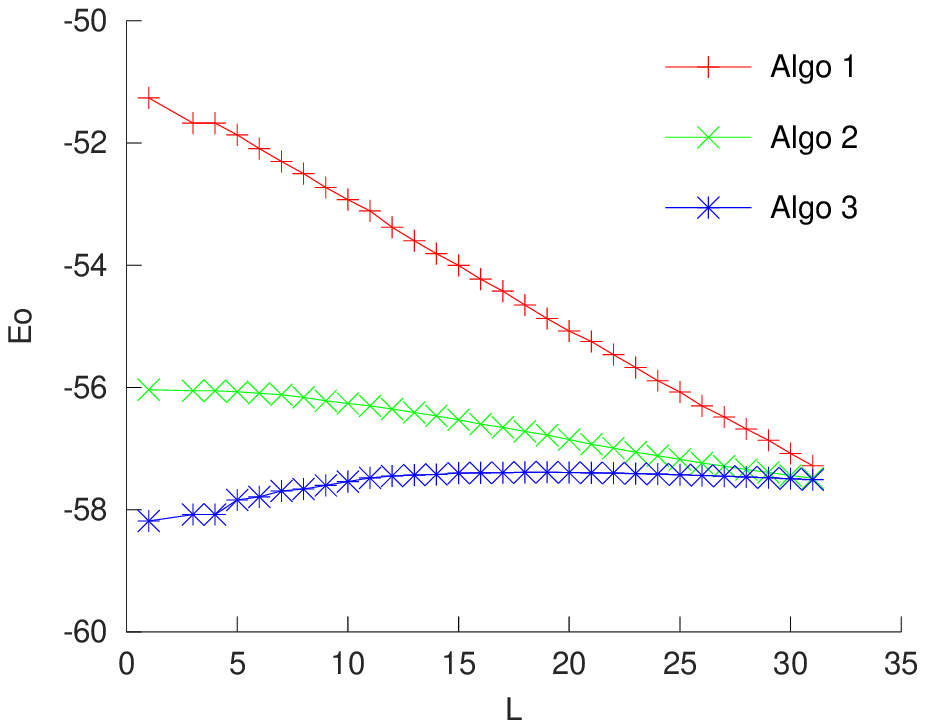}
    \includegraphics[width=6.25cm]{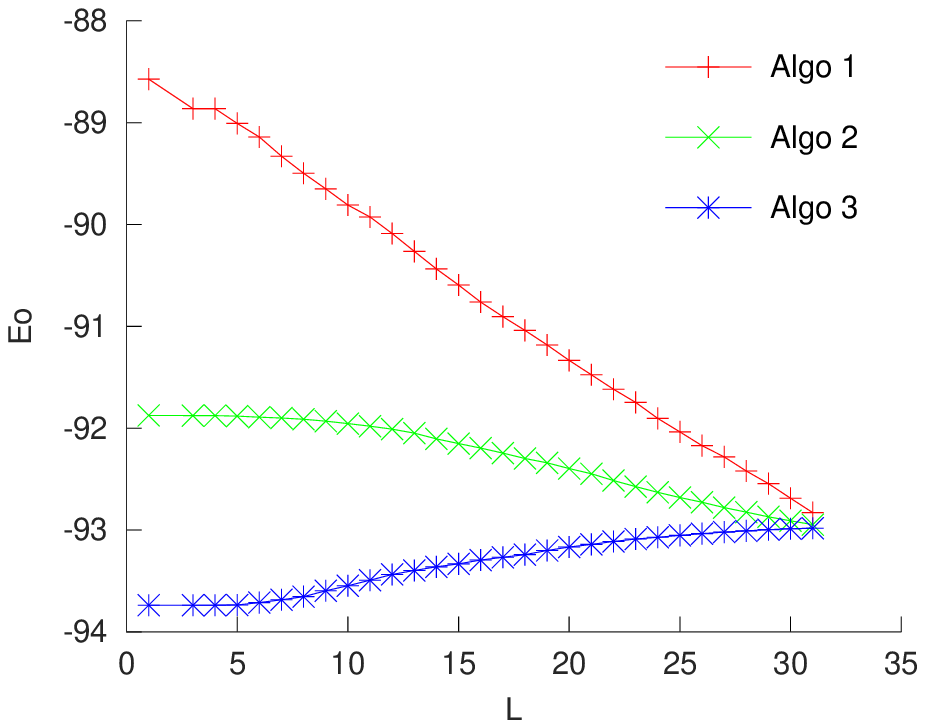}
    \caption{Average ground state energy  of a random Ising chain of 32 spins
      as estimated by the three variants of the MPO-RG algorithm
      versus the number of remaining sites $L$ during the RG process.
      The system is in the disordered Griffiths phase ($h_i\in [1;2]$)
      on the left and in the paramagnetic phase ($h_i\in [2.5;3]$) in the right.
    }
  \label{fig7}
  \end{center}
\end{figure}

The results are presented on figures~\ref{fig3} to \ref{fig7}. The average
ground state energy is plotted versus the number of remaining sites $L$ during
the RG process for the three MPO-RG algorithms. All points from $L=31$ (after
the first renormalization step) to $L=4$ were computed by applying the DMRG
algorithm to the renormalized MPO. The last point $L=1$ corresponds to the
average ground state energy given by the MPO-RG algorithm at the end of the
renormalization, i.e. when there is only one site left. Since the first point
on the right corresponds to the energy after only one renormalization step,
its value is therefore close to the exact value. The figures show a
monotonous evolution with $L$ of the estimates of the ground state energy.
However, a jump is sometimes observed for Algo 1 and 3 at the end of the
calculation, i.e. $L$ small. It seems therefore safer to stop the calculation
at $L\ge 5$ and compute exactly the quantum averages rather than pursuing
the renormalization up to $L=1$.

As can be seen on the figures, the original MPO algorithm (Algo 1) induces
much larger systematic deviations of the ground state energy than the two
other algorithms. In the paramagnetic phase, the systematic deviation grows
approximatively linearly with the number of RG steps, i.e. each iteration
is followed by the same shift of the ground state energy. The relative
deviation at the end of the calculation is about $4.7\%$. In the ferromagnetic
phase, the deviation tends to be larger at the beginning of the renormalization
process. Almost no shift is observed in the last iterations. Nevertheless, the
relative deviation of the ground state energy at the end of the
calculation is about $3.8\%$.

Despite a small modification with respect to Algo. 1, the algorithm with
an improved choice of the block to be renormalized (Algo. 2) turns out to
be surprisingly much more efficient. As can be seen on figures~\ref{fig3}
to \ref{fig7}, the average ground state energy displays a much smaller shift
as the renormalization is performed. The total deviation of the ground
state energy at the end of the calculation is about $1.2\%$ in the
paramagnetic phase and $0.8\%$ in the ferromagnetic phase.

The MPO-RG algorithm with effective interactions (Algo. 3) brings some
improvements with respect to the two other algorithms. In the ferromagnetic
phase, the average ground state energy is systematically lower during the
renormalization process and the relative deviation at the end of the
calculation is about $0.5\%$. However, in the paramagnetic phase,
the average ground state energy goes below the exact one and the relative
deviation is about $-0.8\%$, i.e. the same deviation as Algo. 2 but with
a different sign~\footnote{
Note that DMRG is a variational approach, which therefore guarantees that
the estimated ground-state energy is always higher than the exact one.
In contrast, algo. 3 relies on a perturbative expansion. Therefore, nothing
prevents an energy lower than the ground state energy from being measured.}

\subsection{Stability of the algorithms with more states per block}
In this section, the three variants of the MPO-RG algorithm are compared
for a larger lattice of 240 sites and with 4,8,16 or 32 states per
block during renormalization. The ground state energy is computed at
the end of the renormalization of the chain, i.e. when only one site is
left. It is averaged over 1000 disorder configurations.

  \begin{table}
  \caption{Average ground state energies of a random Ising chain of 240 spins
    as estimated by the three variants of the MPO-RG algorithm for different
    numbers of states kept during the truncation of the Hilbert space.
  }
  \begin{tabular}{llllll}
    \mr
    & $h\in[0.3;0.4]$ & $h\in[0.5;1]$ & $h\in[0.5;2]$ & $h\in[1;2]$ & $h\in[2.5;3]$ \\
    \br
Algo 1 & & & & & \\
4 states
& $-3.00643.10^2$& $-3.12685.10^2$& $-3.60362.10^2$& $-4.04007.10^2$& $-6.80757.10^2$\\
8 states
& $-3.02988.10^2$& $-3.21175.10^2$& $-3.75499.10^2$& $-4.18228.10^2$& $-6.88960.10^2$\\
16 states
& $-3.04019.10^2$& $-3.24527.10^2$& $-3.82430.10^2$& $-4.23943.10^2$& $-6.93308.10^2$\\
32 states
& $-3.04537.10^2$& $-3.26313.10^2$& $-3.85995.10^2$& $-4.26790.10^2$& $-6.95864.10^2$\\
\mr
Algo 2 & & & & & \\
4 states
& $-3.03551.10^2$& $-3.23612.10^2$& $-3.85713.10^2$& $-4.27185.10^2$& $-6.95643.10^2$\\
8 states
& $-3.04476.10^2$& $-3.26770.10^2$& $-3.89493.10^2$& $-4.30330.10^2$& $-6.97327.10^2$\\
16 states
& $-3.04827.10^2$& $-3.28022.10^2$& $-3.91017.10^2$& $-4.31602.10^2$& $-6.97933.10^2$\\
32 states
& $-3.04980.10^2$& $-3.28684.10^2$& $-3.91891.10^2$& $-4.32286.10^2$& $-6.98192.10^2$\\
\mr
Algo 3 & & & & & \\
4 states
& $-3.03697.10^2$& $-3.24632.10^2$& $-4.01471.10^2$& $-4.40873.10^2$& $-6.99829.10^2$\\
8 states
& $-3.04560.10^2$& $-3.27368.10^2$& $-3.98940.10^2$& $-4.35034.10^2$& $-6.98487.10^2$\\
16 states
& $-3.04877.10^2$& $-3.28406.10^2$& $-3.96255.10^2$& $-4.34810.10^2$& $-6.98376.10^2$\\
32 states
& $-3.05011.10^2$& $-3.28941.10^2$& $-3.94581.10^2$& $-4.34575.10^2$& $-6.98377.10^2$\\
\mr
    \end{tabular}
  \label{table1}
\end{table}

  On table~\ref{table1}, the average ground state energies are presented at the same
  points of the phase diagram as in the previous section. For the three algorithms,
  all estimates evolve monotonously as the number of states per block is increased.
  The energies only decrease for Algo 1 and 2 while they increase for Algo 3 at
  the critical point, in the disordered Griffiths phase and in the paramagnetic phase.
  Nevertheless, the estimates of the three algorithms seem to converge towards
  the same value with a convergence which is faster for Algo 3. Assuming that this
  value is the exact ground state energy, one can notice that, as in section 5.1,
  the estimates of Algo. 2 is systematically higher than this exact energy
  while it is lower for Algo. 3 at the critical point and in the paramagnetic
  phase.
  
\begin{table}
    \caption{Average energy gap between the first excited state and the ground state
      of a random Ising chain of 240 spins as estimated by the three variants of the
      MPO-RG algorithm for different numbers of states kept during the truncation of
      the Hilbert space.
  }
  \begin{tabular}{llllll}
    \mr
    & $h\in[0.3;0.4]$ & $h\in[0.5;1]$ & $h\in[0.5;2]$ & $h\in[1;2]$ & $h\in[2.5;3]$ \\
    \br
   Algo 1 & & & & & \\
4 states
& $ 0.058270$& $ 0.10491$& $ 0.10131$& $ 0.46837$& $ 2.9748$\\
8 states
& $ 0.058662$& $ 0.079809$& $ 0.038335$& $ 0.39790$& $ 2.7150$\\
16 states
& $ 0.057870$& $ 0.081128$& $ 0.029123$& $ 0.33381$& $ 2.6969$\\
32 states
& $ 0.059596$& $ 0.076776$& $ 0.022698$& $ 0.29653$& $ 2.7078$\\
\mr
Algo 2 & & & & & \\
4 states
& $ 0.051948$& $ 0.058287$& $ 0.011563$& $ 0.18225$& $ 2.5420$\\
8 states
& $ 0.051221$& $ 0.052873$& $ 0.012093$& $ 0.18069$& $ 2.5071$\\
16 states
& $ 0.050813$& $ 0.052706$& $ 0.012025$& $ 0.18297$& $ 2.4829$\\
32 states
& $ 0.050232$& $ 0.052737$& $ 0.012001$& $ 0.17774$& $ 2.4633$\\
\mr
Algo 3 & & & & & \\
4 states
& $ 0.051741$& $ 0.066224$& $ 0.012145$& $ 0.18851$& $ 2.4050$\\
8 states
& $ 0.050549$& $ 0.053929$& $ 0.013135$& $ 0.17546$& $ 2.3987$\\
16 states
& $ 0.050494$& $ 0.051849$& $ 0.012938$& $ 0.17560$& $ 2.3797$\\
32 states
& $ 0.050058$& $ 0.051016$& $ 0.013564$& $ 0.16924$& $ 2.3828$\\
\mr
    \end{tabular}
  \label{table2}
\end{table}

On table~\ref{table2}, the average gaps between the first excited state and the
ground state energies are presented. Note that in the ferromagnetic and
ordered Griffiths phases, the gap is due to the energy splitting induced by
the small magnetic field $B$. In contrast to the average ground state energies,
the estimates of the three algorithms do not display any monotonous evolution
with the number of states per block. However, we note that the average gap is
about 6000 times smaller than the ground state energy. The ground state and the
first excited state show the same monotonous evolution with the number of states
and their difference, i.e. the gap, displays a monotonous evolution only when it
is larger than the statistical fluctuations introduced by the average over disorder.
For most of the data in table~\ref{table2}, the improvement due to the increase
of the number of states seems to be smaller than these fluctuations.
  
  \begin{table}
    \caption{Average Binder cumulant $1-\overline{\langle m^4\rangle}/3\overline
      {\langle m^2\rangle}^2$, where $m={1\over L}\sum\sigma_i^z$ is the
      magnetization density, of a random Ising chain of 240 spins
    as estimated by the three variants of the MPO-RG algorithm for different
    numbers of states kept during the truncation of the Hilbert space.
  }
  \begin{tabular}{llllll}
    \mr
    & $h\in[0.3;0.4]$ & $h\in[0.5;1]$ & $h\in[0.5;2]$ & $h\in[1;2]$ & $h\in[2.5;3]$ \\
    \br
Algo 1 & & & & & \\
4 states
& $ 0.68639$& $ 0.76002$& $ 0.95962$& $ 0.98771$& $ 0.99436$\\
8 states
& $ 0.68042$& $ 0.73790$& $ 0.94623$& $ 0.98559$& $ 0.99352$\\
16 states
& $ 0.67784$& $ 0.72771$& $ 0.92263$& $ 0.98401$& $ 0.99311$\\
32 states
& $ 0.67660$& $ 0.72264$& $ 0.91182$& $ 0.98315$& $ 0.99297$\\
\mr
Algo 2 & & & & & \\
4 states
& $ 0.68275$& $ 0.74544$& $ 0.92321$& $ 0.98458$& $ 0.99305$\\
8 states
& $ 0.67862$& $ 0.73082$& $ 0.91096$& $ 0.98295$& $ 0.99297$\\
16 states
& $ 0.67693$& $ 0.72371$& $ 0.90518$& $ 0.98228$& $ 0.99292$\\
32 states
& $ 0.67611$& $ 0.71993$& $ 0.90169$& $ 0.98195$& $ 0.99289$\\
\mr
Algo 3 & & & & & \\
4 states
& $ 0.68276$& $ 0.74601$& $ 0.93128$& $ 0.98488$& $ 0.99310$\\
8 states
& $ 0.67863$& $ 0.73090$& $ 0.91675$& $ 0.98301$& $ 0.99299$\\
16 states
& $ 0.67693$& $ 0.72373$& $ 0.90939$& $ 0.98225$& $ 0.99292$\\
32 states
& $ 0.67611$& $ 0.71996$& $ 0.90403$& $ 0.98193$& $ 0.99289$\\
\mr
    \end{tabular}
  \label{table4}
  \end{table}

  We also computed the average Binder cumulant
  \be U=1-{\overline{\langle m^4\rangle}\over 3\overline {\langle m^2\rangle}^2}\ee
  where $m={1\over L}\sum\sigma_i^z$ is the
  magnetization density. The second and forth moments $\langle m\rangle^2$
  and $\langle m\rangle^4$ were evaluated using the technique introduced in Ref.~\cite{Lin}.
  The data are presented in table~\ref{table4}. Again, for a given number of states, Algo. 1
  displays a larger deviation than the two other algorithms. The largest deviation is
  found at the critical point. Note that the moments involved in the definition of the
  Binder cumulant can be written as the sum over the lattice of two and four-point correlation
  functions. The faster convergence of the Binder cumulant indicates that the estimates
  of these correlations are improved, not only at short distances but also over large
  distances. Indeed, the improvement of the renormalisation of a local operator
  propagates in the lattice exponentially fast with the number of iterations because
  of the tree structure of the calculation. In contrast, a local improvement in the DMRG
  algorithm would propagate linearly.

  \subsection{Efficiency of the different algorithms}
  To compare the efficiency of the three algorithms, the execution times
  for the 2-color Ashkin-Teller model with 8 states per site at $\epsilon=1$
  (to be discussed in the next section) are considered. For the different values
  of the transverse field $h$, the execution time was between $1980s$ and $2476s$
  for Algo. 1, between $2315s$ and $2829s$ for Algo. 2, and between $3386s$ and
  $23203s$ for Algo. 3.
Despite the fact that Algo 1 and Algo. 2 differ only by a different order
in which the local Hamiltonians are renormalized, there is an average CPU
overhead of the order of $15\%$ for Algo. 2. The different order of the
renormalizations leads indeed to a smaller gap at the vicinity of the phase
boundaries (see also Table~\ref{table2}). As a consequence, the numerical determination
of the eigenvalues and eigenvectors using the {\tt arpack} library takes
more CPU time. As expected, Algo 3. is much slower due to the extra operations
performed, in particular the determination of $G_0(z)$ by a first
diagonalization and then a conjugate gradient method. On average, the running
time is roughly the double of that of Algo 1. but, for a few points of the
phase diagram, Algo. 3 can be up to ten times slower than Algo. 1.
  
\section{Phase diagram of the 2 and 3-color random Ashkin-Teller models}
In this section, the $N$-color random quantum Ashkin-Teller chain is considered. The model
consists in $N$ quantum Ising chains in a transverse field coupled by 2 and 4-spin interactions.
The Hamiltonian of the model is
\ba H&=&-\sum_{\alpha=1}^N\Big[\sum_{i=1}^{L-1} J_i\sigma_{\alpha,i}^z\sigma_{\alpha,i+1}^z
  +h\sum_{i=1}^{L} \sigma_{\alpha,i}^x\Big] \nonumber\\
  &&\quad -\sum_{\alpha,\beta<\alpha}\Big[\sum_{i=1}^{L-1} K_i\sigma_{\alpha,i}^z\sigma_{\beta,i}^z
  \sigma_{\alpha,i+1}^z\sigma_{\beta,i+1}^z+g\sum_{i=1}^{L} \sigma_{\alpha,i}^x\sigma_{\beta,i}^x\Big]
\ea
where $\sigma_{\alpha,i}^{x,z}$ are spin-$1/2$ operators. In the following, the case where
$J_i$ and $K_i$ are random couplings is studied. The intra-chain couplings $J_i$ are uniformly
distributed in $[0;1]$ and the ratio $\epsilon=K_i/J_i=g/h$ is kept constant.
The Hamiltonian is cast to a MPO whose matrices read in the bulk of the chain
\be\fl A_i=\pmatrix{ \identite\otimes\identite &
  -\sqrt{J_i}\sigma^z\otimes\identite &
  -\sqrt{J_i}\identite\otimes\sigma^z &
  -\sqrt{K_i}\sigma^z\otimes\sigma^z &
  -h_i(\sigma^x\otimes\identite+\identite\otimes\sigma^x)\atop
  -g\sigma^x\otimes\sigma^x   \cr
  0 & 0 & 0 & 0 & \sqrt{J_i}\sigma^z\otimes\identite\cr
  0 & 0 & 0 & 0 & \sqrt{J_i}\identite\otimes\sigma^z\cr
  0 & 0 & 0 & 0 & \sqrt{K_i}\sigma^z\otimes\sigma^z\cr
  0 & 0 & 0 & 0 & \identite\otimes\identite}\ee
for the $N=2$ color Ashkin-Teller model. In the case $N=3$, the matrices are $8\times 8$.

\subsection{The 2-color random Ashkin-Teller model}
In the pure case, i.e. when $J_i$ and $H_i$ are uniform over the chain, the phase diagram of the
2-color Ashkin-Teller model shows three second-order transition lines merging at a tricritical
point at $K=J$~\cite{Kohmoto}. Two of them belong to the Ising universality class. Along the third
one, the critical exponents depends on $K$. To distinguish the three phases, two order parameters,
magnetization $m$ and polarization $p$, can be defined:
    \be m={1\over L}\sum_{i=1}^L\sigma_{1,i}^z,\hskip 1truecm
    p={1\over L}\sum_{i=1}^L\sigma_{1,i}^z\sigma_{2,i}^z.\ee
In the following, the two Binder cumulants associated to these two order parameters will
be considered:
    \be U_m=1-{\overline{\langle m^4\rangle}\over 3\overline{\langle m^2\rangle}^2},
    \hskip 1truecm
    U_p=1-{\overline{\langle p^4\rangle}\over 3\overline{\langle p^2\rangle}^2}.\ee
In presence of disorder, the phase diagram of the $N=2$ quantum Ashkin-Teller model
is qualitatively unchanged. However, along the three transition lines, the critical behavior
is governed by the same Infinite-Randomness Fixed Point as the random Ising chain in a
transverse field~\cite{Hoyos2}. Only at the tricritical point where these lines meet, a
new Infinite-Randomness Fixed Point is observed.
\\

\begin{figure}
   \begin{center}
	\psfrag{h}[tc][tc][1][0]{$h$}
    \psfrag{Um}[Bc][Bc][1][1]{$U_m$}
	\psfrag{Up}[Bc][Bc][1][1]{$U_p$}
    \includegraphics[width=6.25cm]{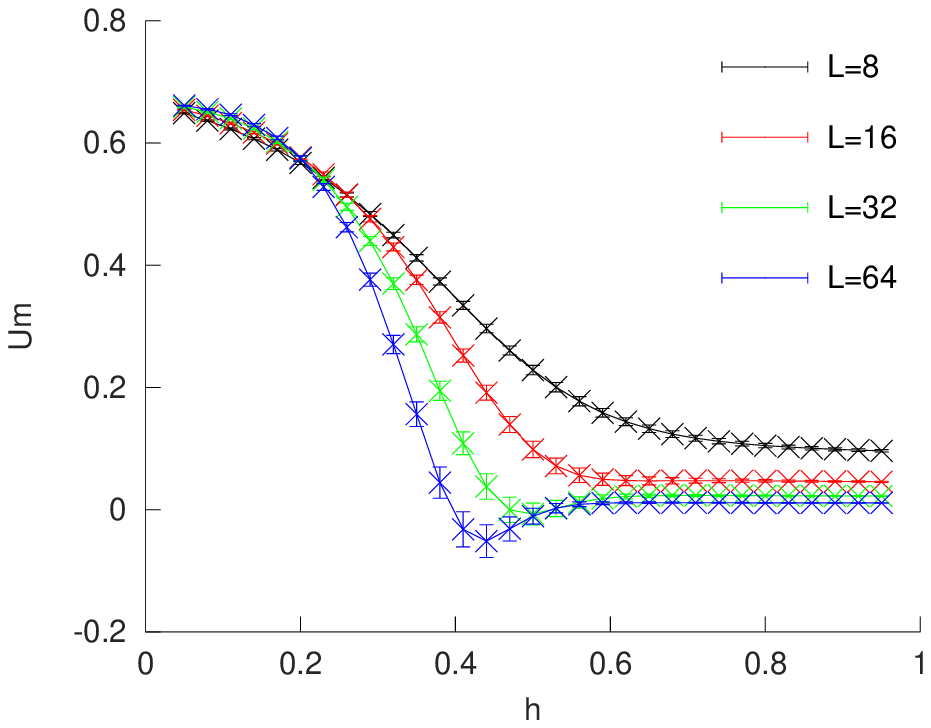}
    \includegraphics[width=6.25cm]{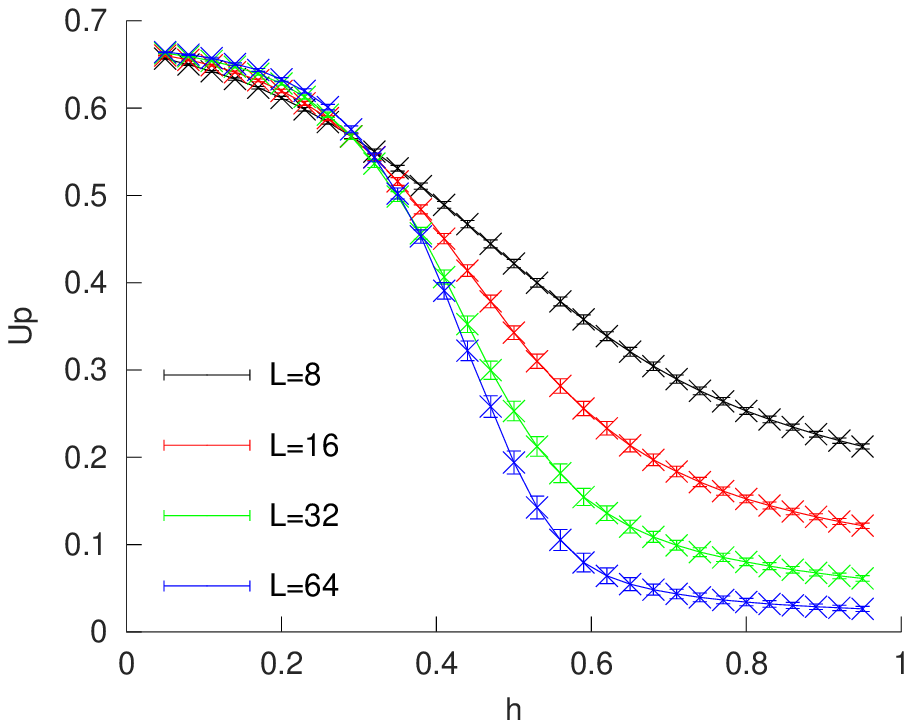}
    \caption{Average Binder cumulant of magnetization (left) and polarization (right)
             for the 2-color Ashkin-Teller model with $\epsilon=2$.
             The data have been computed using Algo. 1 with 4 states
             per site. Error bars correspond to the standard deviation taken over
			the 10.000 disorder configurations.}
    \label{fig22}
  \end{center}
\end{figure}

Algorithm 1 with $4$ states per site, equivalent to the original SDRG algorithm, is not able to give correct Binder cumulants $U_m$ and $U_p$. As can be seen on figure~\ref{fig22} in the particular case $\epsilon=2$, the magnetization cumulant $U_m$ displays a dip and takes negative values. This anomalous behaviour is also observed with $U_p$ at small $\epsilon$. Moreover, figure~\ref{fig22} shows that the crossings of $U_m$ and $U_p$ occur at two critical transverse fields $h_c$ that are close to each other. This contradicts the fact that for $\epsilon>1$, two distinct second-order phase transitions are expected. The critical lines, determined from the crossing of the curves associated to two successive lattice sizes, are not monotonous and therefore cannot be considered as reliable. Keeping 8 states per site instead of 4 slightly improves the shape of the curves. A dip is still present but is smaller. For $\epsilon>1$, two distinct transition lines are now observed. The cumulant $U_p$ leads to a rather well-defined transition line but with estimates of $h_c$ still much too small compared to other algorithms. For $U_m$, crossings can be found only for the smallest lattice sizes but not for the largest ones.
\\

\begin{figure}
\begin{center}
                     \psfrag{h}[tc][tc][1][0]{$h$}
                        \psfrag{Um}[Bc][Bc][1][1]{$U_m$}
                        \psfrag{Up}[Bc][Bc][1][1]{$U_p$}
    \includegraphics[width=6.25cm]{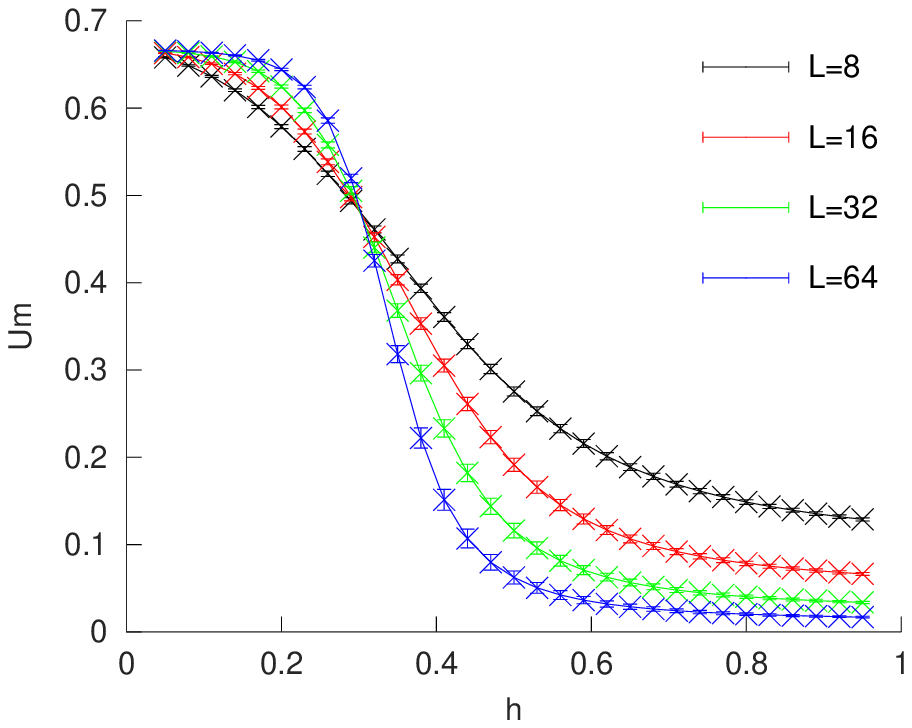}
    \includegraphics[width=6.25cm]{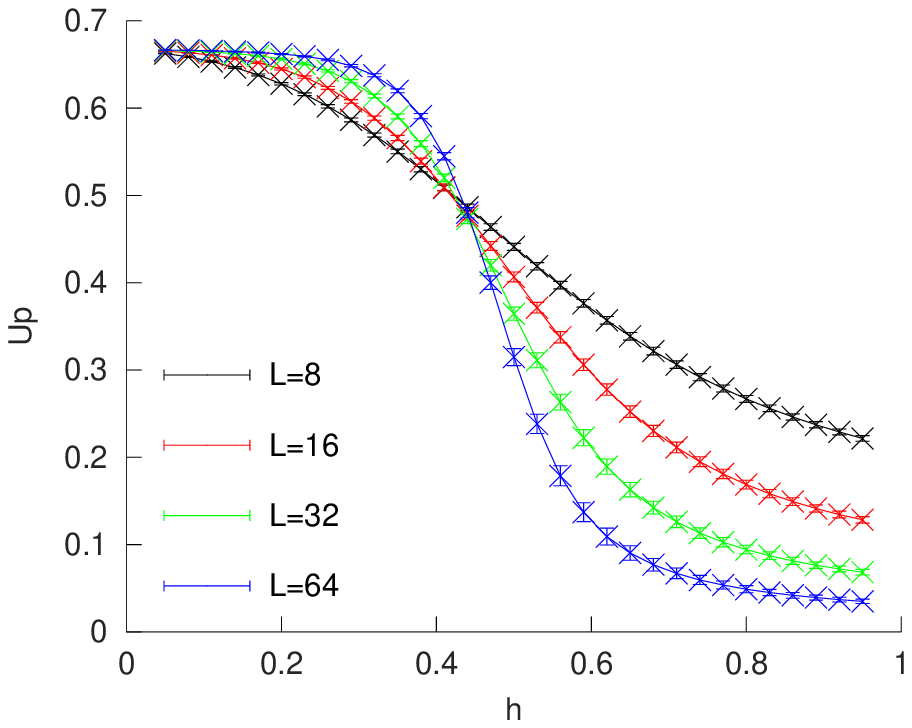}
    \caption{Average Binder cumulant of magnetization (left) and polarization (right)
             for the 2-color Ashkin-Teller model with $\epsilon=2$.
             The data have been computed using Algo. 2 with 4 states
             per site. Error bars correspond to the standard deviation taken over
			 the 10.000 disorder configurations.}
    \label{fig42}
  \end{center}
\end{figure}

       \begin{figure}
                \begin{center}
                        \psfrag{epsilon}[tc][tc][1][0]{$\epsilon$}
                        \psfrag{hc}[Bc][Bc][1][1]{$h_c$}
    \includegraphics[width=8cm]{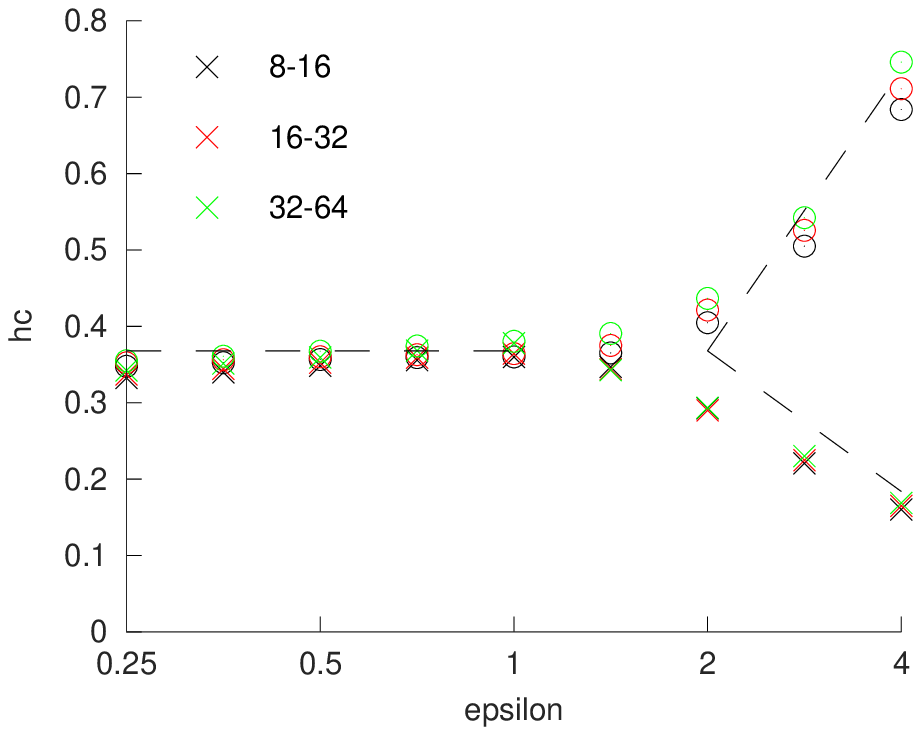}
    \caption{Phase diagram of the 2-color Ashkin-Teller model 
             obtained from the crossings of the Binder cumulants $U_m$ (crosses) 
             and $U_p$ (circles). The colors are associated to the pair
             of lattice sizes (see the legend) used to find the crossing of the cumulants.
             The data have been computed using Algo. 2 with 8 states per site.
             The dashed lines correspond to the self-dual line $h=e^{-1}$ at $\epsilon\le 1$
             and the two branches $h=\epsilon/2e$ and $h=2/\epsilon e$ predicted by SDRG
             and assumed to be exact in the limit $\epsilon\rightarrow +\infty$~\cite{Hoyos2}.
    }
    \label{fig109}
  \end{center}
\end{figure}

Even though Algo. 2 consists only in a simple change of the order in which the sites
are decimated, the improvement for the Binder cumulants is drastic. As shown on figure~\ref{fig42}, no dip is present anymore. The phase diagram is greatly improved and
is consistent with what is expected (figure~\ref{fig109}). For $\epsilon\le 1$, the critical transverse field is close to the value $h_c=e^{-1}\simeq 0.37$ imposed by self-duality. Keeping 8 states per site instead of 4 leads to a small improvement of the critical fields $h_c(\epsilon)$ for the largest lattice sizes. Using Algo. 3 leads to another small improvement of the critical fields. The data is very close for the two algorithms using the same number of states per site.
\\

As can be seen on figure~\ref{fig109}, the largest lattice sizes lead to a better agreement with the
self-dual line. However, at the tricritical point $\varepsilon=1$, the largest
lattice sizes ($L=32-64$) goes slightly above the expected value $h_c=e^{-1}$.
In the regime $\epsilon>1$, the numerical data has reached the SDRG
predictions at $\epsilon\simeq 4$. Again, the largest lattice sizes go
beyond these SDRG predictions in in the upper branch.
There are two possible explanations for this deviation at the largest
lattice sizes: the number of disorder realisations, kept equal to 1000,
becomes too small at large lattice sizes to sample correctly the rare events
or the number of states, kept equal to 8 for all lattice sizes, should
be increased with the lattice size to reproduce with the same fidelity the
ground state of the system. 

\subsection{The 3-color random Ashkin-Teller model}
For $N\ge 3$, the pure $N$-color quantum Ashkin-Teller chain undergoes a single first-order phase transition. It is well-known that, in classical systems, first-order transitions are softened by randomness through a mechanism uncovered by Imry and Wortis~\cite{ImryWortis}. For two-dimensional classical systems, the Aizenmann-Wehr theorem states that an infinitesimal amount of disorder is sufficient to make the transition continuous~\cite{HuiBerker,HuiBerker2,Aizenman,Aizenman2}. Goswani {\sl et al.} argued that the same occurs in the quantum case~\cite{Goswani}. Analyzing the SDRG flow equations, they showed that a small coupling between random Ising chains is an irrelevant perturbation at the infinite-disorder fixed point. This implies that the critical behavior of the random 3-color Ashkin-Teller model is the same as the one of the random Ising chain in a transverse field, in contrast to what was observed in the classical case~\cite{Bellafard,Bellafard2}. A numerical iteration of the SDRG rules
confirmed this statement and extended the conclusion to the strong coupling
regime~\cite{Hoyos1,Hoyos3}. In the meantime, the Aizenmann-Wehr theorem has been
generalized to quantum systems~\cite{Greenblatt,Greenblatt2}.
\\

        \begin{figure}
                \begin{center}
                        \psfrag{h}[tc][tc][1][0]{$h$}
                        \psfrag{Um}[Bc][Bc][1][1]{$U_m$}
                        \psfrag{Up}[Bc][Bc][1][1]{$U_p$}
    \includegraphics[width=6.25cm]{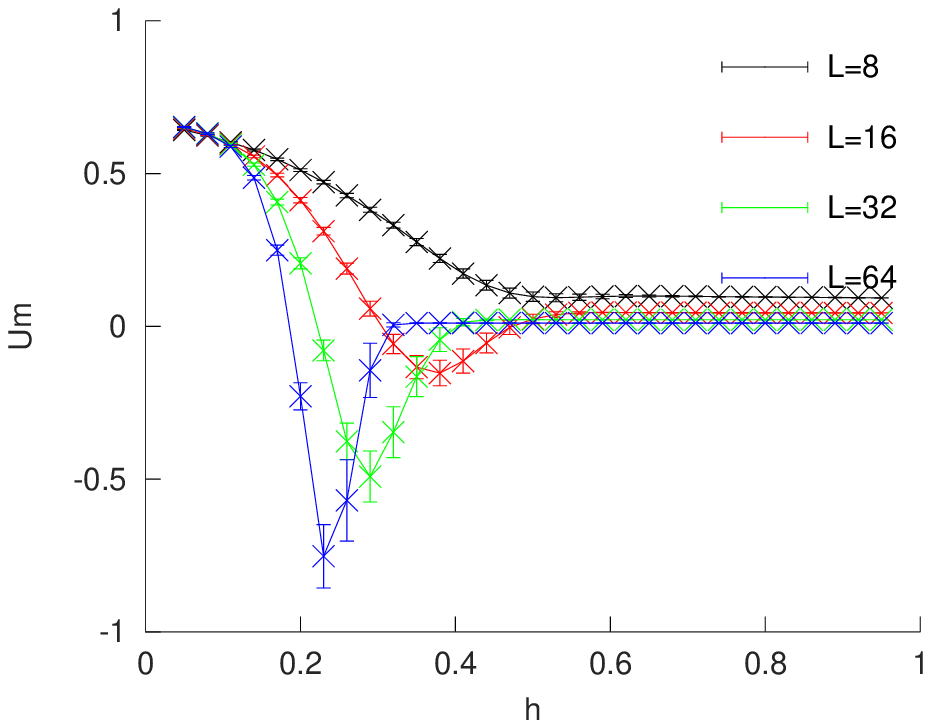}
    \includegraphics[width=6.25cm]{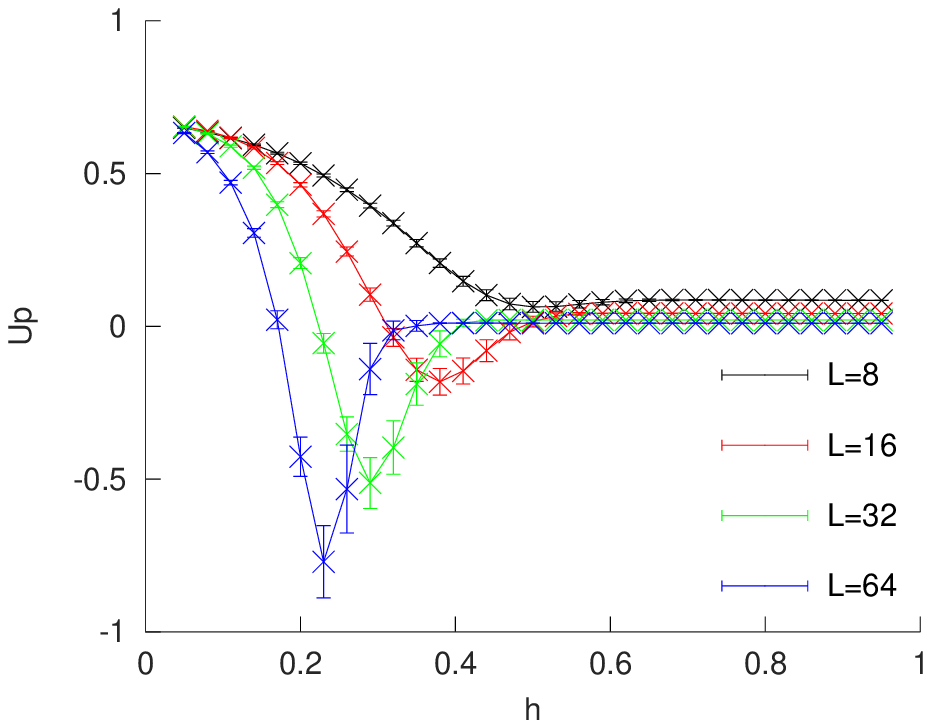}
    \caption{Average Binder cumulant of magnetization (left) and polarization (right)
             for the 3-color Ashkin-Teller model with $\epsilon=2$.
             The data have been computed using Algo. 1 with 8 states
             per site. Error bars correspond to the standard deviation taken over
			 the 10.000 disorder configurations.}
    \label{fig142}
  \end{center}
\end{figure}

Using Algo. 1 with 8 states, equivalent to the original SDRG algorithm, both magnetization
and polarization cumulants $U_m$ and $U_p$ display a dip that becomes deeper as $\epsilon$ is increased. The example of $\epsilon=2$ is presented on Figure~\ref{fig142}. The shape of the curves and the fact that the dip becomes deeper when the lattice size increases is typical of a first-order phase transition and therefore contradicts the results of the litterature. Keeping 16 states per site instead of 8 leads to very similar Binder cumulants.

\begin{figure}
	\begin{center}
	\psfrag{h}[tc][tc][1][0]{$h$}
    \psfrag{Um}[Bc][Bc][1][1]{$U_m$}
                        \psfrag{Up}[Bc][Bc][1][1]{$U_p$}
    \includegraphics[width=6.25cm]{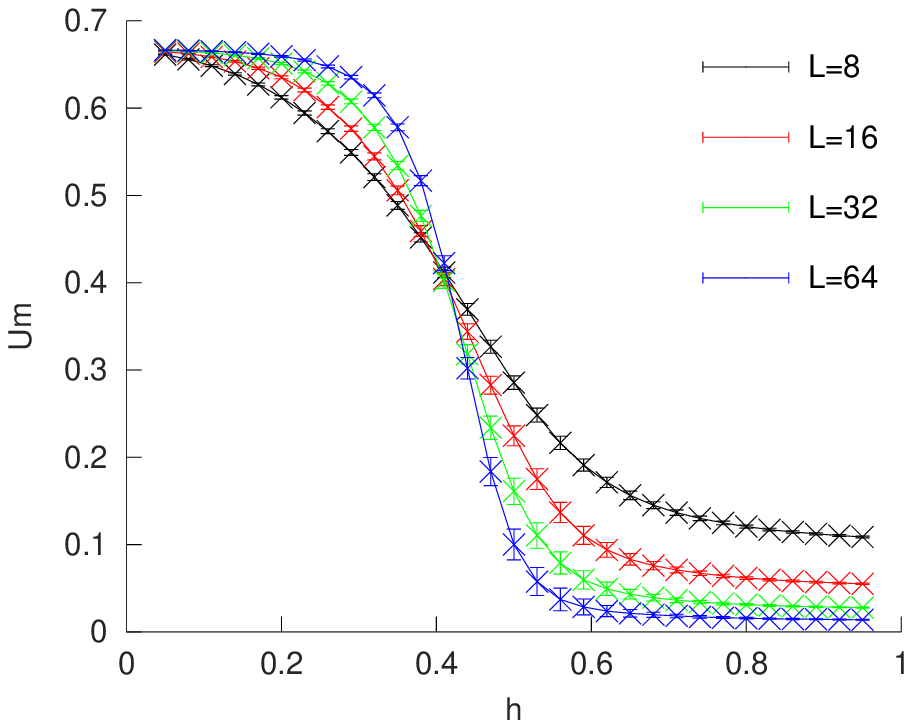}
    \includegraphics[width=6.25cm]{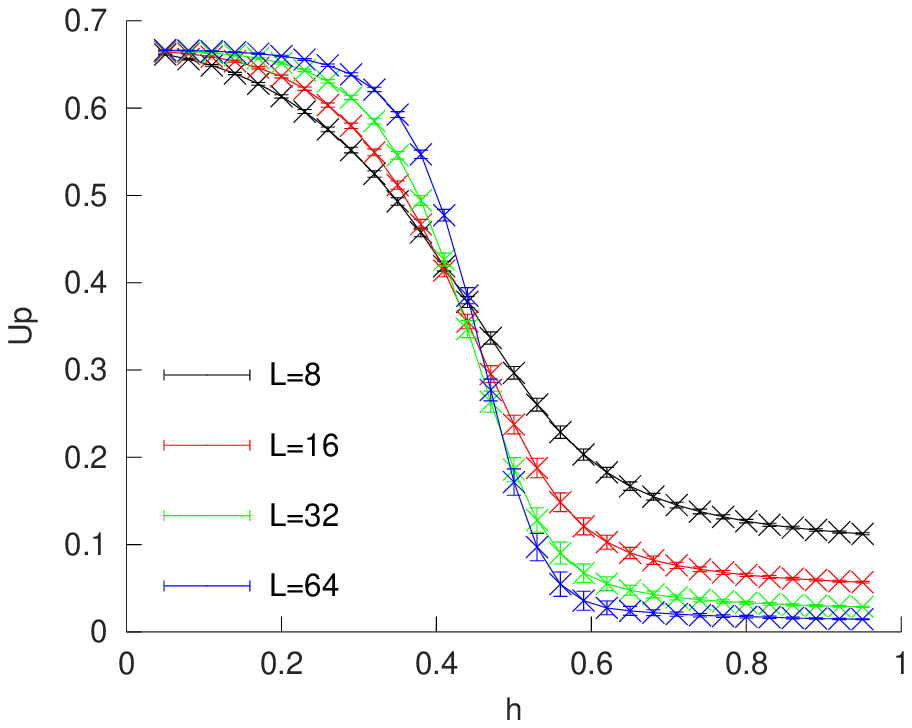}
    \caption{Average Binder cumulant of magnetization (left) and polarization (right)
             for the 3-color Ashkin-Teller model with $\epsilon=2$.
             The data have been computed using Algo. 2 with 8 states
             per site. Error bars correspond to the standard deviation taken over
			 the 10.000 disorder configurations.}
    \label{fig162}
  \end{center}
\end{figure}

\begin{figure}
                \begin{center}
                        \psfrag{epsilon}[tc][tc][1][0]{$\epsilon$}
                        \psfrag{hc}[Bc][Bc][1][1]{$h_c$}
    \includegraphics[width=8cm]{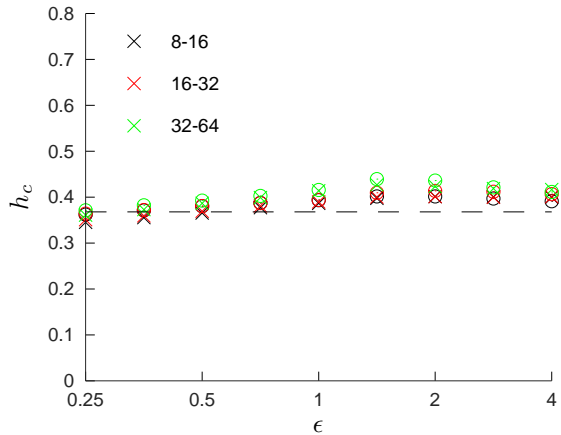}
    \caption{Phase diagram of the 3-color Ashkin-Teller model 
             obtained from the crossings of the Binder cumulants $U_m$ (crosses) 
             and $U_p$ (circles). The colors are associated to the pair
             of lattice sizes (see the legend) used to find the crossing of the cumulants.
             The data have been obtained using Algo. 2 with 16 states per site.
             The dashed line corresponds to the self-dual line $h=e^{-1}$.
    }
    \label{fig229}
  \end{center}
\end{figure}

Using now Algo. 2 or 3 with 8 states, very different results are obtained as shown on figure~\ref{fig162}. The shape is now typical of a continuous phase transition and the critical field can be estimated from the crossings of the curves for two successive lattice sizes. However, for $\epsilon\ge 2\sqrt 2$, the decay of the polarization Binder cumulant $U_p$ is slightly too slow at strong transverse fields. As a consequence, the crossings of the Binder cumulant $U_p$ is shifted to larger transverse fields. Keeping 16 states instead of 8 leads to well-behaved curves. It appears that the Binder cumulants $U_m$ and $U_p$ display crossings at the same transverse fields, for both $\epsilon\le 1$ and $\epsilon>1$. This confirms the existence of a unique phase transition, and therefore the absence of a partially ordered phase, as already proposed in Ref.~\cite{Hoyos3} based on the analysis of the RG flow. The critical transverse field remains close to the self-dual value $h_c=e^{-1}\simeq 0.37$ (figure~\ref{fig229}). However, the largest lattice sizes display
the largest deviation to this self-dual field. A largest number of disorder configurations or of states kept
during the renormalization should improve the accuracy.

\section*{Conclusions}
We have presented two variants of the MPO renormalization algorithm.
In the first one (Algo 2), the choice of the blocks to be merged
and renormalized takes into account the couplings with the neighboring
blocks of the chain. The renormalization differs therefore only
by the order in which the blocks are grouped together. Nevertheless,
it is observed that this simple modification improves the accuracy
of the ground state energy by a factor at least 4 in all regions
of the phase diagram of the random Ising chain in a transverse field.
In the second algorithm (Algo 3), effective interactions are generated
to take into account the highest eigenstates to be discarded during
the renormalization. We observe a small improvement of the accuracy
of the ground state energy. However, in contrast to Algo 1 and 2,
this algorithm gives smaller estimates of the ground state energy
than the exact one in the paramagnetic and disordered Griffiths phases.
The smallest energy is therefore not the necessarily the best one in
this case. We note that the algorithm may be improved by taking into
account the three-site interaction, as well as higher orders in the
Dyson expansion. Finally, it was shown that the two algorithms are
stable as the number of states kept during the renormalization is
increased.
\\

These new algorithms have been applied to the random 2 and 3-color
Ashkin-Teller models. Unlike the original MPO renormalization algorithm,
they are shown to give well-behaved magnetization and polarization Binder
cumulants from which the phase diagram can be reconstructed.
Since the Binder cumulant involves
second and forth-order moments, that can be written as the sum over the
lattice of two and four-point correlation functions, the drastic improvement
brought by the proposed algorithms shows that this improvement is not only
local but extends to long-distance correlation functions. In contrast to DMRG,
any local improvement is indeed spread exponentially fast over the lattice
by the tree-like structure of the calculation.
As expected, the phase diagram of the 2-color Ashkin-Teller model is
qualitatively unchanged in presence of disorder while the first-order phase
transition of the 3-color Ashkin-Teller model becomes continuous.
The technique is however limited to relatively small lattice sizes:
at large lattice sizes, small systematic deviations of the phase boundaries
from the expected ones have indeed been observed. These deviations can possibly
be reduced by either increasing the number of states kept during renormalization
or the number of disorder realisations.
\\

Recently, a different route, based on entanglement renormalization~\cite{Vidal},
has been investigated to improve the accuracy of MPO-RG~\cite{Goldsborough2}.
The ground-state is constructed as a tensor network involving not
only unitaries but also disentanglers. The computational effort
is however increased with the number of variational parameters. We note that
the structure of the tensor network is determined by first applying
SDRG to the random chain. The results of the present paper show that
the accuracy could probably be greatly improved in a simple way by replacing
SDRG by Algo 2. It would be therefore very interesting to investigate the use of
Algo 2 to construct the tensor network for entanglement renormalization.

\end{document}